\documentclass[letterpaper, 10 pt, conference]{ieeeconf}  % Comment this line out if you need a4paper

\IEEEoverridecommandlockouts                              % This command is only needed if 
\usepackage{graphics} % for pdf, bitmapped graphics files
\usepackage{epsfig} % for postscript graphics files
\usepackage{amsmath}   % For advanced math commands
\usepackage{amssymb}   % Additional math symbols
\usepackage{mathtools} % Enhancements for math formatting
\usepackage{siunitx}  % Handles numerical and scientific formatting

\usepackage{graphicx}
\usepackage{tikz}
\usepackage{tabularx} 
\usetikzlibrary{patterns,shapes,arrows,fit,positioning,shapes.geometric,shapes.symbols,shapes.misc,decorations,calc,backgrounds}
\usepackage{pgfplots}
\usepackage{pgfmath}

\usepackage{amssymb}
\pgfplotsset{compat=newest}
\pgfplotsset{plot coordinates/math parser=false}
\usepackage{subcaption}
\usepackage{stfloats}
\usepackage{float}
\usepackage{caption}
\usepackage{makecell}
\usepackage{tikz, pgfplots}
\usetikzlibrary{positioning}
\usetikzlibrary{shapes.geometric}
\usetikzlibrary{shapes.symbols}
\usetikzlibrary{decorations.pathreplacing}
\usepackage{lipsum}
\usetikzlibrary{calc,positioning,shapes.geometric,shapes.symbols,shapes.misc}
\usepackage{array}
\usepackage{multirow, makecell}
\usepackage{hyperref}
\usepackage{tabularx} % also loads 'array' package
\usepackage[export]{adjustbox}  

\newlength\figurewidth
\newlength\figureheight

\usepackage{tikz}
\usepackage{tikz-cd}
\usepackage{pgfplots}
\pgfplotsset{compat=1.14}

\title{\LARGE \bf
Robust EEG Functional Connectivity Metrics for Decoding Action Observation Conditions and Observed Actions
}

\author{Anh T. Nguyen, \textit{Member, IEEE}$^{1}$, Zachary A. Rentala$^{2}$,  and Michelle J. Johnson, \textit{Member, IEEE}$^{3}$
\thanks{*This work was supported by the Department of Physical Medicine and Rehabilitation at the University of Pennsylvania and by the Eunice Kennedy Shriver National Institute of Child Health \& Human Development of the National Institutes of Health (NIH) under Award Number F31HD102165, the Vingroup Scholarship Program for Master’s and Ph.D. Degrees Overseas Study, and the American Heart Association (AHA) Predoctoral Fellowship under Award Number 25PRE1372607. The content does not necessarily represent the views of the NIH.}% <-this % stops a space
\thanks{$^{1}$Anh T. Nguyen is with the School of Science and Applied Science, Department of Bioengineering, University of Pennsylvania, Philadelphia, PA, USA
        {\tt\small tuna28ng@seas.upenn.edu}}
\thanks{$^{2}$Zachary A. Rentala is with the College of Arts and Sciences, Department of Cognitive Science, University of Pennsylvania, Philadelphia, PA, USA
    {\tt\small zrentala@sas.upenn.edu}}%
\thanks{$^{3}$Dr. Michelle J. Johnson is an Associate Professor with the Department of Physical Medicine and Rehabilitation and BioEngineering. She directs the Rehab Robotics Lab (A GRASP Lab), University of Pennsylvania, Philadelphia, PA, USA{\tt\small johnmic@pennmedicine.upenn.edu}}
}

\begin{document}
\maketitle
\thispagestyle{empty}
\pagestyle{empty}
%%%%%%%%%%%%%%%%%%%%%%%%%%%%%%%%%%%%%%%%%%%%%%%%%%%%%%%%%%%%%%%%%%%%
\begin{abstract}
Action observation (AO) paradigms probe motor‑system engagement, yet the electroencephalographic (EEG) functional‑connectivity (FC) metrics that best capture AO dynamics remain uncertain. This pilot study benchmarks five sensor‑level FC metrics---coherence (COH), imaginary coherence (iCOH), phase‑locking value (PLV), partial directed coherence (PDC) and spectral Granger causality (SpcG)---for decoding AO stimuli in five healthy adults.
EEG was recorded while participants viewed upper‑limb actions performed by human or robot agents and non‑action controls. Ten motor‑area channels were analyzed in the alpha (8–12 Hz) and beta (13–30 Hz) bands. Trial‑wise 10 $\times$ 10 FC matrices were supplied to different classifiers to solve two tasks: (i) six‑class AO‑condition decoding and (ii) five‑class action‑type decoding.
Across both tasks, volume‐conduction–invariant metrics consistently outperformed their counterparts: iCOH yielded the highest macro‐AUC across most classifiers, with PDC and SpcG closely following.  Graph neural networks (GNNs) delivered the most robust and stable performance across all FC metrics, with convolutional neural network (CNNs) and random forests ranking a close second. These results underscore the importance of FC features that suppress zero‐phase‐lag and capture directed interactions, and demonstrate GNNs’ unique ability to exploit the inherent graph structure, offering clear guidance for metric and model selection in future, larger‐scale studies of action‐observation–related brain activity.  

\end{abstract}

%%%%%%%%%%%%%%%%%%%%%%%%%%%%%%%%%%%%%%%%%%%%%%%%%%%%%%%%%%%%%%%%%%%%%%%%%%%%%%%%
\section{INTRODUCTION}

Understanding how the brain processes observed actions is critical for advancing applications in brain-computer interfaces, cognitive neuroscience, and human-robot interaction. EEG-based decoding of action observation (AO) offers a noninvasive window into the underlying neural mechanisms, enabling the study of functional brain connectivity during passive observation of others’ movements. While AO paradigms are often motivated by their potential in neurorehabilitation \cite{borges2022action}, this study focuses on decoding AO-related brain activity in healthy individuals to evaluate the performance and generalizability of signal processing pipelines.

Functional connectivity (FC)—which characterizes statistical dependencies between brain regions—has shown promise as a feature representation for decoding cognitive and sensorimotor processes from EEG \cite{shahhosseini2022functional}. FC metrics can be grouped into non-directed (e.g., coherence, phase locking value) and directed (e.g., spectral Granger causality, partial directed coherence) measures, each capturing different aspects of neural coupling \cite{bastos2016tutorial}. Importantly, these metrics differ in their susceptibility to confounding factors such as volume conduction, which can artificially inflate connectivity estimates due to instantaneous field spread. Despite growing use of FC in AO and motor decoding tasks, it remains unclear which FC metrics are robust when measuring the neural response to multiple visual stimuli.

Recent work has shown that machine learning classifiers, including support vector machines (SVMs), random forests, and deep neural networks such as convolutional (CNNs) and graph-based (GNNs) models, can be trained on FC-derived features for accurate EEG classification in clinical and cognitive settings \cite{kleplEEGBasedGraphNeural2022, maratova2022comparative, musaeus2019oscillatory}. However, most studies focus on a single classification task, which limits insight into model and feature generalizability. In contrast, our study evaluates classification across two label sets derived from the same experimental EEG data: (1) AO condition decoding (e.g., observing a robot or human agent, on the left or right side), and (2) observed action type decoding (e.g., waving, punching, or arm movement). This dual-task design enables a cross-task evaluation of FC metrics and classification models.

By training multiple classifiers on FC matrices computed from various metrics, we aim to identify which approaches consistently yield high decoding accuracy across both tasks. 
Importantly, directed functional connectivity metrics (e.g., PDC and SpcG) were included in our hypotheses because they provide causal estimates of information flow between cortical regions. As highlighted by Zhou et al. \cite{zhouComparisonDirectedFunctional2022}, directed functional connectivity measures can capture causal information flow within motor-related networks, offering a more precise characterization of action observation–related dynamics than undirected indices. Thus, we anticipated that volume-conduction–invariant and directed metrics would provide complementary robustness in decoding AO-related brain activity. 
This work contributes to the identification of robust EEG-based decoding strategies for diverse cognitive state inference tasks.

%%%%%%%%%%%%%%%%%%%%%%%%%%%%%%%%%%%%%%%%%%%%%%%%%%%%%%%%%%%%%%%%%%%%%%%%%%%%%%%%
\section{METHODS}

\begin{figure*}[hb]
    \centering
        \valign{%
          #\cr
          \hbox{\subcaptionbox* {(A)}[.25\linewidth]{%
            \includegraphics[width=\linewidth]{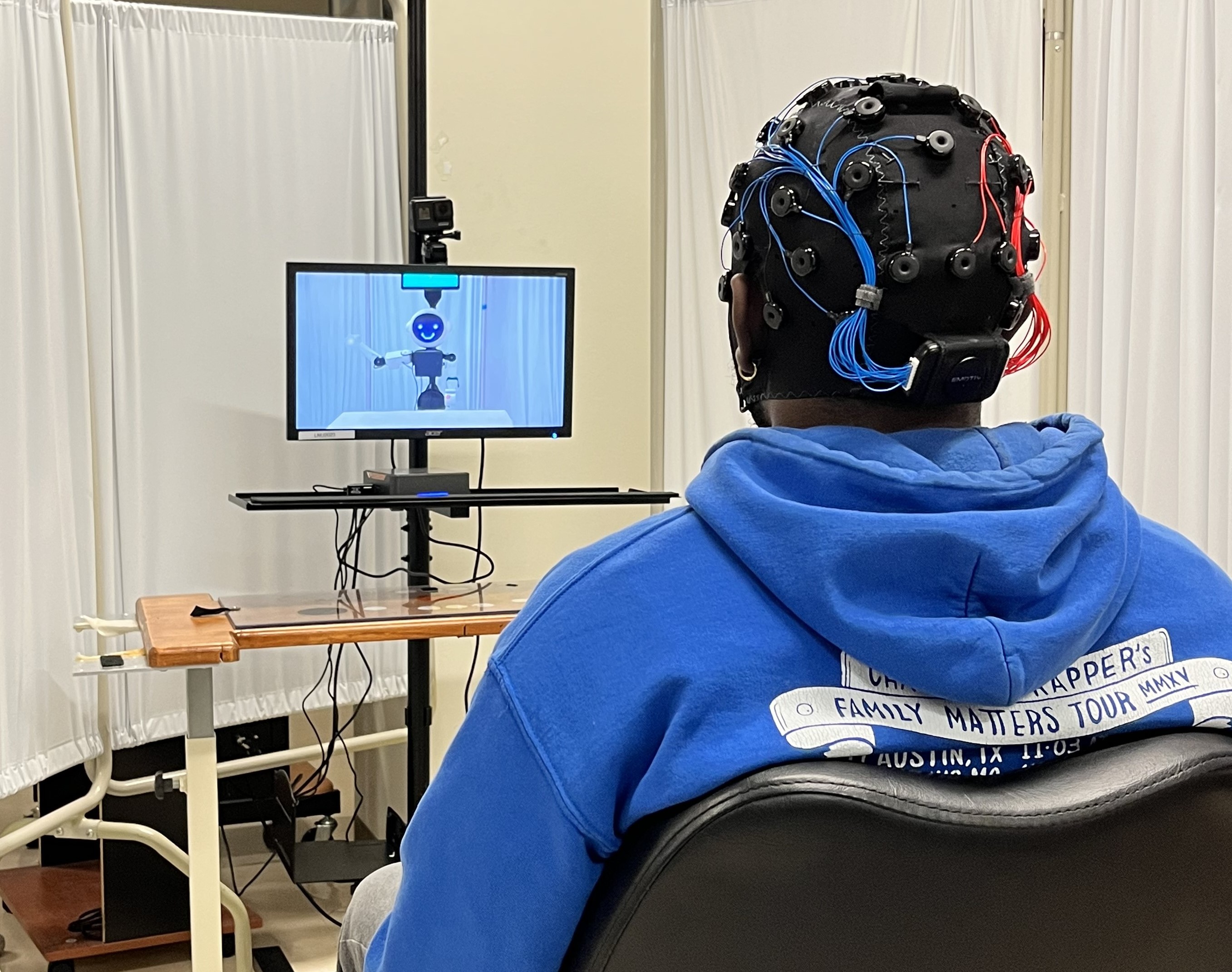}%
          }}
          \vfill \centering
          \hbox{\subcaptionbox*{(B)}[.25\linewidth]{%
            \includegraphics[width=\linewidth]{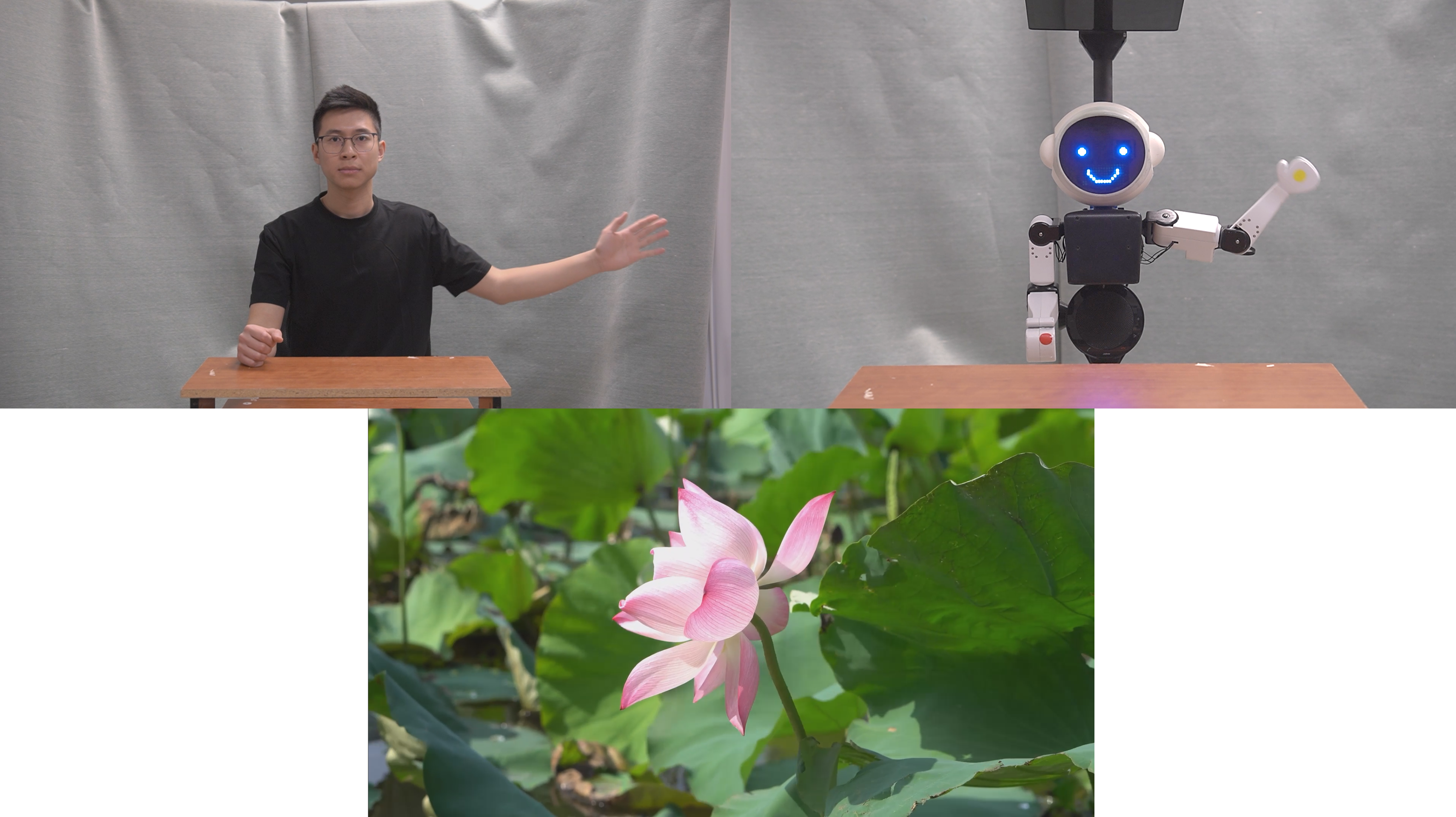}%
          }}
          \cr
          % \noalign{\hfill}
          % \hbox{\subcaptionbox*{(C)}[.115\linewidth]{%
            % \includegraphics[width=\linewidth]{images/flo-gen-shot-squared.jpg}%
          % }}\cr
          \cr\noalign{\hfill}
          \hbox{\subcaptionbox*{(C)}[.7\linewidth]{%
            \includegraphics[width=\linewidth]{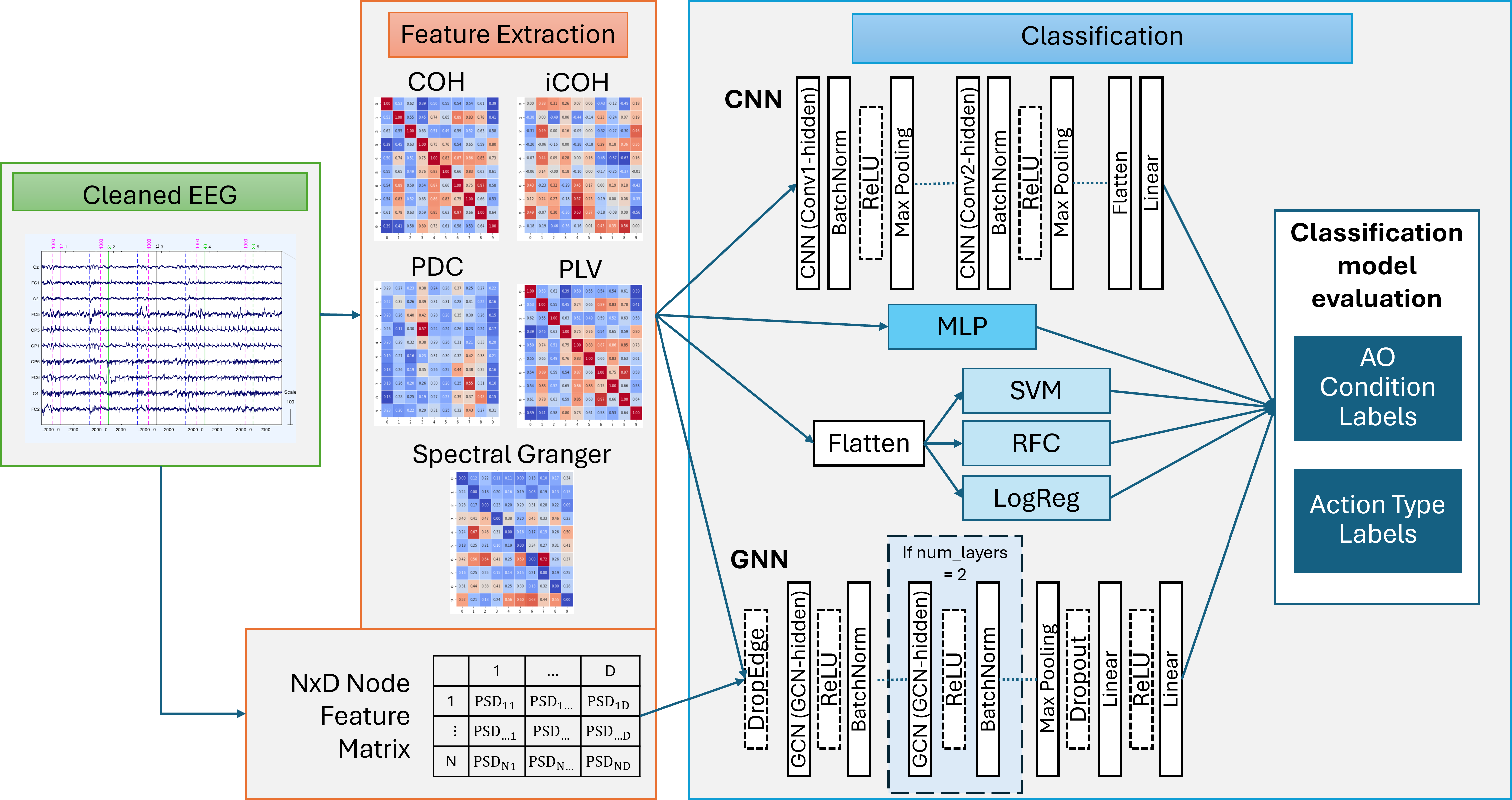}%
          }}\cr  
        }
    \caption{(A) One of the subjects participating in the experiment, observing a video clip of "robot" condition (B) Examples of different AO conditions: human actor, robot actor, and control (landscape videos) (C) EEG-Based Functional Connectivity Classification Pipeline: EEG data are preprocessed, segmented into segments, and analyzed using COH, iCOH, PDC, PLV, and SpcG methods. The resulting FC adjacency matrices are processed by GNN, CNN, SVM, RF, and LogReg classifiers to predict six AO experiment conditions. }\label{images}
\end{figure*}

\subsection{Experimental Setup}

Five healthy, right-handed individuals (aged 21–29 years) with no history of neurological or psychiatric disorders participated in the action observation (AO) experiment while their EEG data was recorded. This study was approved by the University of Pennsylvania Institutional Review Board (IRB No. 830126) on June 29, 2021.
Participants observed a randomized sequence of video stimuli depicting unimanual upper limb actions performed by either a humanoid social robot, Flo \cite{sobrepera2021design}, or one of two human actors (one male, one female). They were seated 150 cm away from a 24-inch LCD screen (Fig. \ref{images}A) and instructed to remain still with their hands on their lap while watching the videos.

The AO stimuli consisted of 5000-ms videos, each showcasing one of five distinct upper-limb actions: air punching, backward-forward arm swing, lateral arm swing, overhead arm raise, or waving. Each action was presented from an allocentric viewpoint and performed using both left and right arms by all three agents (robot and two humans), resulting in four experimental AO conditions: human-left, human-right, robot-left, and robot-right (Fig. \ref{images}B). Landscape videos featuring natural scenes were included as a visual control condition, and all trials began with a 1000-ms fixation cross to establish a pre-stimulus EEG baseline.
The AO task was structured in a counterbalanced block design using the PsychoPy software package \cite{peirce2019psychopy2} to minimize order effects. Each participant completed 120 trials in total. These trials were used to define two complementary classification tasks:

\begin{itemize}
    \item AO condition decoding: Classifying the observed agent and direction of motion (i.e., human-left, human-right, robot-left, robot-right) as well as control conditions (baseline and landscape).
    \item Action-type decoding: Classifying the specific upper-limb action being observed.
\end{itemize}

Both decoding tasks were derived from the same EEG recordings and experimental structure, enabling direct comparison of model and feature robustness across label types. In total, six experimental conditions were annotated per trial: baseline, landscape, and the four AO conditions. Action labels were independently assigned based on the motion depicted in the video clip.

\subsection{EEG Acquisition and Preprocessing}
The Emotiv EPOC Flex EEG system (Emotiv Inc., San Francisco, U.S.A.), equipped with 32 Ag-AgCl electrodes, was used to collect EEG data \cite{williams2020validation}. Electrodes were positioned according to the international 10–20 system, with two common-mode sensors placed at the left and right mastoids for referencing. EEG signals were digitized at 1,024 Hz, processed using a 5th-order digital sine filter, and subsequently downsampled to 128 Hz. 

Following data collection, EEG preprocessing was performed offline using MATLAB (The MathWorks, Inc., Natick, MA, USA) and EEGLAB toolbox \cite{delorme2004eeglab}. The raw EEG data underwent bandpass filtering (1–40 Hz) and artifact removal, where visually identified noisy channels were eliminated. The data were then re-referenced to the average, and further cleaned using independent component analysis, allowing for the identification and removal of artifacts such as eye blinks, muscle activity, and cardiac signals \cite{lee1998independent}. The cleaned EEG signals were epoched into trials, spanning from the onset of the fixation cross to the end of the subsequent video stimulus. Segments exceeding 100 µV in amplitude were discarded to eliminate artifacts such as eye blinks, muscle movements, or electrical noise.
The cleaned datasets were then split into individual trials and organized by condition. One-second fixation cross segments were used as the baseline condition, while for the other conditions (landscape, human-left, human-right, robot-left, and robot-right), segments from 1 to 3 seconds post-video onset were extracted and used for FC analysis.

\subsection{EEG Functional Connectivity}

To compute functional connectivity (FC), we selected ten EEG channels previously associated with AO-related cortical activity \cite{sun2021functional}: FC1, C3, FC5, CP5, CP1, CP6, FC6, C4, FC2, and Cz. EEG signals were bandpass-filtered into the alpha (8–12 Hz) and beta (13–30 Hz) bands using the multitaper spectral estimation method \cite{li2022multitaper}, which improves frequency resolution and reduces spectral leakage. FC metrics were computed between each channel pair, producing 10$\times$10 adjacency matrices for each trial and band. Metrics were calculated using the FieldTrip toolbox \cite{oostenveld2011fieldtrip} and averaged across each frequency band. 

We computed five widely used FC metrics: coherence (COH), imaginary coherence (iCOH), phase locking value (PLV), partial directed coherence (PDC), and spectral Granger causality (SpcG).

\begin{itemize}
    \item COH measures frequency-domain phase synchrony, normalized to [0, 1]. It is sensitive to volume conduction effects due to inclusion of zero-lag components \cite{spironelli2013beta}.
    \item iCOH isolates the imaginary component of coherency, explicitly removing instantaneous (zero-phase) interactions, making it robust to volume conduction \cite{chiarion2023connectivity}.
    \item PLV computes the consistency of phase differences across trials using amplitude-normalized signals. Like COH, it is undirected and phase-based.
    \item PDC estimates directed influences via normalized Fourier-transformed coefficients from a multivariate autoregressive model, enabling frequency-resolved Granger causality \cite{baccala2001partial}.
    \item SpcG also quantifies directional causal relationships in the frequency domain, based on power-normalized contributions of predictors in an MVAR model \cite{hu2012causality}.
\end{itemize}

COH, PLV, and iCOH are non-directed metrics, while PDC and SpcG provide directed estimates. Among these, iCOH and PDC are considered more robust against volume conduction. All metrics were applied separately to alpha and beta bands for comparative analysis.

\subsection{Classifiers}
Calculating the FC values produced 10$\times$10 adjacency matrices that capture pairwise connectivity between selected EEG electrodes. These matrices served as input features for two classification tasks: (1) decoding action observation (AO) conditions, and (2) decoding observed action types. To evaluate the discriminative utility of each FC metric, we trained six classifiers: three traditional machine learning models—Support Vector Machine (SVM), Logistic Regression (LogReg), and Random Forest (RF)—and three deep learning models—Multilayer Perceptron (MLP), Convolutional Neural Network (CNN), and Graph Neural Network (GNN). Each classifier was applied to the FC matrices across both decoding tasks, and performance was compared across FC metrics and model types.

\subsubsection{Convolutional Neural Network}
CNN is a deep learning model designed for processing grid-structured data, widely used in image classification \cite{ajrashallownetworks}. It consists of convolutional blocks, where each convolutional layer applies a kernel (filter) to extract spatial features\cite{yamashita2018convolutional}. Batch normalization stabilizes training by normalizing activations, while the rectified linear unit (ReLU) activation function introduces non-linearity to improve feature learning. A max pooling layer then reduces feature map size, lowering computational costs while retaining important spatial structures \cite{yamashita2018convolutional}.

The batch size (\texttt{batch\_sz}) hyperparameter determines the number of samples processed per iteration, while hidden dimension (\texttt{hidden}) defines the size of the fully connected layers. Kernel sizes (\texttt{ksz\_0} and \texttt{ksz\_1}) describes the dimensions of the filter, while \texttt{padding} ensures spatial dimensions remain consistent. The learning rate (\texttt{lr}) controls weight updates during optimization, balancing convergence speed and stability. This study’s CNN architecture was adapted from \cite{ajrashallownetworks}, which successfully classified Alzheimer’s patients, individuals with frontotemporal dementia, and healthy participants.

\subsubsection{Graph Neural Network}
GNN is a deep learning framework designed to analyze structured data represented as graphs, where nodes correspond to EEG electrodes and edges encode connectivity strengths defined by an adjacency matrix. We implemented a Graph Convolutional Network (GCN) inspired by Klepl et al.'s architecture \cite{kleplEEGBasedGraphNeural2022}. The model takes as input a graph $ G = \{N, E, F\} $, where $ N $ denotes the set of nodes, $ E $ represents the edge connections, and $ F $ corresponds to the node feature set. Each node feature vector was derived from the Power Spectral Density (PSD), computed in 1 Hz increments from 4 to 30 Hz, yielding a 27-dimensional feature vector per electrode.  

The GCN model architecture is built upon a message-passing framework, where node features are iteratively refined through interactions with their local neighborhood. The model consists of 1-3 graph convolutional layers (\texttt{n\_layers}), each applying spectral filtering to capture spatial dependencies among the EEG electrodes. In a single GCN layer, the representation of node $ i $ at layer $ l $ is updated based on its own features and the maximum-weighted influence from its neighboring nodes:  

\begin{equation} 
x_{i}^{l} = \Theta_{1} x_{i}^{l-1} + \Theta_{2} \max_{j\in G_{i}} e_{i j} x_{j}^{l-1} 
\end{equation}  

where $ \Theta_1 $ and $ \Theta_2 $ are trainable weight matrices, and $ e_{ij} $ represents the edge weight between node $ i $ and its neighbor $ j $. This localized feature aggregation enables the model to capture spatial relationships among EEG electrodes efficiently.  

Following the graph convolutional layers, a global max pooling operation is applied to consolidate node-level embeddings into a single graph-level representation, which are then processed by two fully connected layers. The first fully connected layer reduces the feature dimensionality, while the final classification layer employs log-softmax activation to predict the output class.  
To improve model generalization and mitigate overfitting, dropout regularization (\texttt{dropout}) is applied at both the node and edge levels. Specifically, edge dropout (\texttt{drop\_edge})  is incorporated to randomly remove a fraction of graph connections during training, preventing the model from over-relying on specific structural relationships. The GNN also uses the \texttt{batch\_sz}, \texttt{hidden}, and \texttt{lr} hyperparameters.

\subsubsection{Multilayer Perceptron}
MLP is a modern feed-forward neural network \cite{abd2023multi}. This study's MLP model consists of one to three hidden layers, which is controlled by the \texttt{n\_layers} hyperparameter. Between each layer is an ReLU activation function. MLP model also uses the \texttt{batch\_sz}, \texttt{hidden}, and \texttt{lr} hyperparameters.

\subsubsection{Support Vector Machine}
An SVM maximizes the margins of a hyperplane separator in multidimensional space in order to optimize the generalizability of the model \cite{wang2017research}. This study's SVM was implemented with the hyperparameters \texttt{cost} and \texttt{gamma} which represent the strength of regularization and the inverse size of the decision boundary respectively. 

\subsubsection{Logistic Regression}
LogReg generalizes a binomial logistic regression to problems with multiple classes. It does so by solving multiple one-vs-all classifications where the class with the highest score is chosen. LogReg has a single hyperparameter: \texttt{cost}, which controls the strength of regularization by penalizing large weights in the model \cite{lavalley2008logistic}.
% -- lower values encourage stronger regularization, helping to prevent overfitting, while higher values allow the model to fit the training data more closely 

\subsubsection{Random Forest Classifier}
RF algorithm builds an ensemble of decision trees into a "forest." It is based on the bagging method which assumes that combining learning model increases the accuracy and stability of the classifier \cite{balabied2023utilizing}. RF is controlled by two hyperparameters: \texttt{n\_est} and \texttt{max\_dep} which control the number of trees and the maximum number of decisions that can be made respectively.

\begin{table}[b]
\caption{Macro-AUC of Models Across Different FC Measures – AO Condition Classification}
\label{table:AUC_all_model_FC_AO_cond}
\vspace{-1em}
\begin{center}
\resizebox{\linewidth}{!}{%
\begin{tabular}{c c c c c c}
\hline
\hline
 & iCOH & COH & PDC & PLV & SpcG \\
\hline
CNN    & \textbf{0.995 $\pm$ 0.007} & 0.812 $\pm$ 0.036          & \textbf{0.996 $\pm$ 0.006} & 0.809 $\pm$ 0.037          & \textbf{0.992 $\pm$ 0.009} \\
GNN    & \textbf{0.997 $\pm$ 0.004} & \textbf{0.995 $\pm$ 0.008} & \textbf{0.995 $\pm$ 0.008} & \textbf{0.990 $\pm$ 0.013} & \textbf{0.994 $\pm$ 0.009} \\
MLP    & \textbf{0.985 $\pm$ 0.011} & 0.816 $\pm$ 0.035          & \textbf{0.986 $\pm$ 0.012} & 0.816 $\pm$ 0.037          & \textbf{0.913 $\pm$ 0.024} \\
LogReg & 0.585 $\pm$ 0.045          & 0.816 $\pm$ 0.036          & 0.857 $\pm$ 0.031          & 0.817 $\pm$ 0.034          & 0.842 $\pm$ 0.033          \\
RF     & \textbf{0.992 $\pm$ 0.009} & 0.805 $\pm$ 0.036          & \textbf{0.999 $\pm$ 0.002} & 0.811 $\pm$ 0.037          & \textbf{0.999 $\pm$ 0.001} \\
SVM    & \textbf{0.996 $\pm$ 0.007} & 0.823 $\pm$ 0.034          & \textbf{0.943 $\pm$ 0.021} & 0.815 $\pm$ 0.033          & 0.851 $\pm$ 0.032          \\
\hline
\hline
\end{tabular}%
}
\end{center}
% \vspace{-2em}
\end{table}

\begin{table}[b]
\caption{Macro-AUC of Models Across Different FC Measures – Action Type Classification}
\label{table:AUC_all_model_FC_action_type}
\vspace{-1em}
\begin{center}
\resizebox{\linewidth}{!}{%
\begin{tabular}{c c c c c c}
\hline
\hline
 & iCOH & COH & PDC & PLV & SpcG \\
\hline
CNN    & \textbf{0.990 $\pm$ 0.022} & 0.519 $\pm$ 0.080          & \textbf{0.991 $\pm$ 0.023} & 0.526 $\pm$ 0.079          & \textbf{0.990 $\pm$ 0.022} \\
GNN    & \textbf{1.000 $\pm$ 0.000} & \textbf{1.000 $\pm$ 0.000} & \textbf{1.000 $\pm$ 0.000} & \textbf{1.000 $\pm$ 0.003} & \textbf{1.000 $\pm$ 0.000} \\
MLP    & \textbf{0.969 $\pm$ 0.042} & 0.520 $\pm$ 0.080          & 0.896 $\pm$ 0.055          & 0.516 $\pm$ 0.080          & 0.890 $\pm$ 0.057          \\
LogReg & 0.604 $\pm$ 0.079          & 0.496 $\pm$ 0.086          & 0.656 $\pm$ 0.085          & 0.499 $\pm$ 0.082          & 0.626 $\pm$ 0.086          \\
RF     & \textbf{0.998 $\pm$ 0.011} & 0.513 $\pm$ 0.077          & \textbf{0.998 $\pm$ 0.007} & 0.513 $\pm$ 0.085          & \textbf{0.999 $\pm$ 0.004} \\
SVM    & \textbf{0.996 $\pm$ 0.017} & 0.527 $\pm$ 0.082          & 0.868 $\pm$ 0.059          & 0.527 $\pm$ 0.086          & 0.640 $\pm$ 0.085          \\
\hline
\hline
\end{tabular}%
}
\end{center}
\vspace{-2em}
\end{table}

\subsection{Implementation of Model Training and Evaluation}

% Model training and evaluation were implemented in Python using the PyTorch and Scikit-Learn libraries. 
Deep learning models (MLP, CNN, GNN) were implemented using the PyTorch and were trained using the cross-entropy loss function.
% , defined as:
% \begin{equation*}
% \mathcal{L}_{\text{CE}} = - \sum_{o=1}^{B} \sum_{c=1}^{C} y_{o,c} \log(p_{o,c})
% \end{equation*}
% where \(C\) is the number of classes, \(B\) is the batch size, \(y_{o,c}\) is the ground truth one-hot encoding for sample \(o\), and \(p_{o,c}\) is the predicted probability for class \(c\).
Optimization for neural network models was performed using the Adam optimizer, which combines momentum and adaptive learning rate updates. Traditional machine learning models (SVM, LogReg, RF) were trained using Scikit-Learn implementations, which internally apply convex optimization or ensemble learning.
Model selection and evaluation followed a stratified 15-fold cross-validation procedure (\(k=15\)) to ensure balanced class representation. For each fold, training was repeated 5 times to account for variance due to model initialization. The final reported performance for each classifier and FC metric reflects the average across all repetitions and folds.

Hyperparameter tuning was conducted using Optuna \cite{akiba2019optuna}, which employs Bayesian optimization with a Tree-structured Parzen Estimator (TPE) to efficiently explore the hyperparameter search space, specifically:
% \begin{itemize}
%   \item \texttt{batch\_sz}$\in\{8,16,32\}$
%   \item \texttt{n\_layers}$\in\mathbb{Z}\cap[1,3]$
%   \item \texttt{hidden}$\in\{32,64,128\}$
%   \item \texttt{lr}$\in\{0.001,0.005,0.01\}$
%   \item \texttt{ksz\_0}$\in\{3,5\}$
%   \item \texttt{ksz\_1}$\in\{1,2\}$
%   \item \texttt{padding}$\in\{1,2,4\}$
%   \item \texttt{cost}$\in\mathbb{R}\cap[0.001,10]$
%   \item \texttt{gamma}$\in\mathbb{R}\cap[0.001,1]$
%   \item \texttt{n\_est}$\in\mathbb{Z}\cap[10,200]$
%   \item \texttt{max\_dep}$\in\mathbb{Z}\cap[2,20]$
%   \item \texttt{dropout}$\in[0.1,0.5]$
%   \item \texttt{drop\_edge}$\in[0.1,0.5]$
% \end{itemize}
\texttt{batch\_sz} $\in\{8,16,32\}$, \texttt{n\_layers} $\in\mathbb{Z}\cap[1,3]$, \texttt{hidden} $\in\{32,64,128\}$, \texttt{lr} $\in\{0.001,0.005,0.01\}$, \texttt{ksz\_0} $\in\{3,5\}$, \texttt{ksz\_1} $\in\{1,2\}$, \texttt{padding} $\in\{1,2,4\}$, \texttt{cost} $\in\mathbb{R}\cap[0.001,10]$, \texttt{gamma} $\in\mathbb{R}\cap[0.001,1]$, \texttt{n\_est} $\in\mathbb{Z}\cap[10,200]$, \texttt{max\_dep} $\in\mathbb{Z}\cap[2,20]$, \texttt{dropout} $\in[0.1,0.5]$, \texttt{drop\_edge} $\in[0.1,0.5]$.
Optimization objectives included macro-averaged area under the ROC curve (macro-AUC) and balanced accuracy, computed from one-versus-all classification tasks to evaluate each model's ability to distinguish between classes. 
Training was run for a maximum of 300 epochs with early stopping enabled: training terminated if the validation loss did not improve for 15 consecutive epochs.
% Table \ref{table:possible_hyper_params} presents the hyperparameters of the best models which was determined by ranking AUC.

For final model selection, classifiers were evaluated using multiple metrics, including balanced accuracy, macro-averaged precision, recall, F1-score. 
% Macro-AUC served as the primary ranking criterion due to its robustness in imbalanced multi-class settings, while balanced accuracy and macro F1 score provided additional insight into classification fairness. 
To assess model stability, 50 repeated cross-validation runs were performed, and performance variability was quantified using the maximum deviation between the mean and the 5th and 95th percentiles—yielding conservative, distribution-free error estimates.

\begin{figure}[b]
    \centering
    \begin{subfigure}[t]{\linewidth}
        \includegraphics[width=\linewidth]{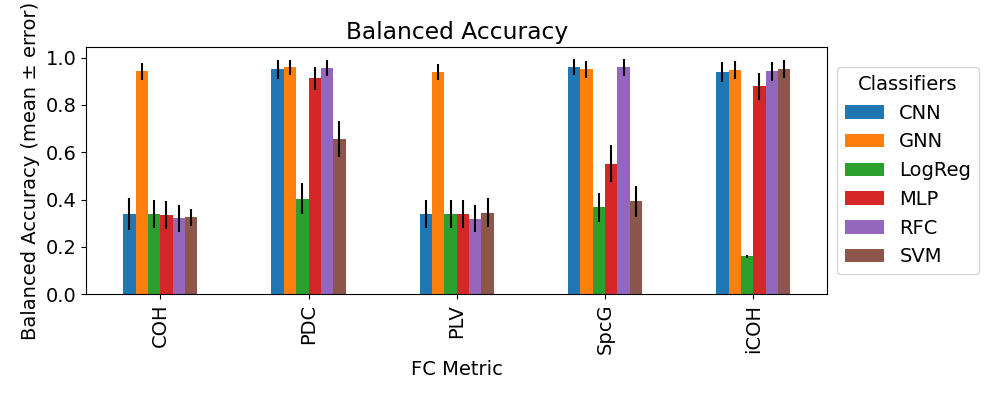}
        \caption{ }
        \label{fig:balanced_accuracy_all_models_bar_AO_cond}
    \end{subfigure}

    % \vspace{-1em}

    \begin{subfigure}[t]{\linewidth}
        \includegraphics[width=\linewidth]{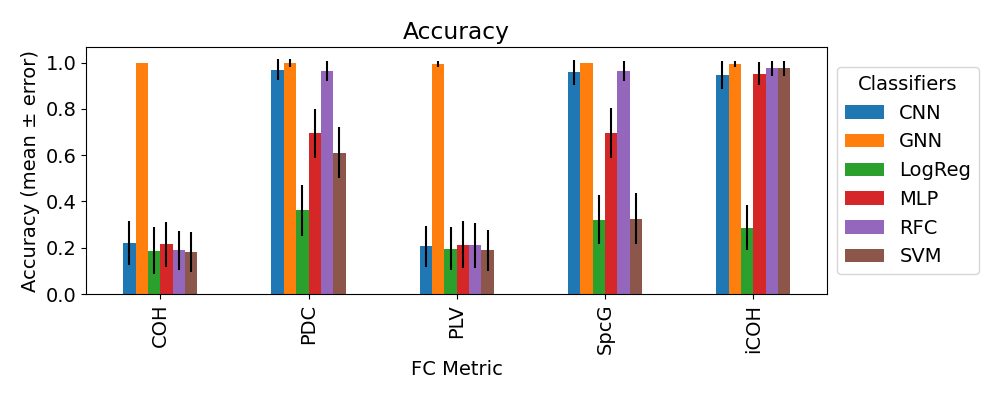}
        \caption{ }
        \label{fig:accuracy_all_models_bar_action_type}
    \end{subfigure}

    \caption{Classifier accuracy for (a) AO condition and (b) action type decoding across FC metrics; error bars represent max deviation between the mean and the 5th/95th percentiles.}
    \label{fig:performance_bars}
    \vspace{-2em}
\end{figure}

\section{RESULTS}

The AO‐condition decoding task comprised approximately 115 baseline trials and 25 trials for each non‐baseline condition. To assess class imbalance, we computed the imbalance ratio (IR), obtaining a mean IR of 4.37 and a maximum IR of 5.95, indicating that the most frequent class had nearly six times more samples than the least frequent. The action‐type decoding task, by contrast, was inherently balanced.

\subsection{Functional Connectivity Metrics}

Tables~\ref{table:AUC_all_model_FC_AO_cond} and \ref{table:AUC_all_model_FC_action_type} report the mean macro‐AUC (± maximum deviation between the mean and the 5th/95th percentiles) for each classifier across the five FC metrics in the AO‐condition and action‐type tasks, respectively. In both tasks, imaginary coherence (iCOH) achieved the highest macro‐AUC in five of the six classifiers (AO: up to 0.997\(\pm\)0.004; action type: up to 1.000\(\pm\)0.000). Directed metrics PDC and SpcG also yielded strong performance, with macro‐AUC \(>\) 0.90 in five AO‐condition models and in three action‐type models. By comparison, non‐directed metrics coherence (COH) and phase‐locking value (PLV) produced macro‐AUC \(>\) 0.80 across all AO‐condition models (COH: up to 0.995\(\pm\)0.008; PLV: up to 0.990\(\pm\)0.013), but underperformed in the action‐type task (mean macro‐AUC \(\approx\) 0.50 for all but the GNN).

Figure~\ref{fig:performance_bars} presents balanced accuracy for each classifier–metric combination in the AO‐condition task (correcting for class skew) and overall accuracy in the action‐type task. Balanced accuracy, precision, recall, and F$_1$‐score showed the same rank ordering across FC metrics as macro‐AUC.
% , confirming the robustness of iCOH, PDC, and SpcG as feature representations.

\subsection{Classifier Performance}

Across both decoding tasks, the graph neural network (GNN) achieved uniformly high macro‐AUC (\(>\) 0.99) for all FC metrics, and perfect overall accuracy (1.000\(\pm\)0.000) in the action‐type task. Convolutional neural networks (CNNs) and random forests (RFs) followed closely on the top‐performing metrics (iCOH, PDC, SpcG), with CNN macro‐AUC \(\geq\) 0.990 in AO‐condition decoding and RF macro‐AUC \(\geq\) 0.992. Multilayer perceptrons (MLPs) and support vector machines (SVMs) demonstrated more variable performance, particularly on weaker features (COH, PLV). Logistic regression (LogReg) was the poorest performer, with macro‐AUC ranging from 0.585 to 0.857 in AO‐condition decoding and 0.496 to 0.626 in action‐type decoding.

\begin{figure*}[!t]
  %------------------- first panel -------------------%
  \begin{subfigure}[b]{\textwidth}
    \centering
    % \begin{adjustbox}{clip,trim=0 0 0.1\linewidth 0}%
      \includegraphics[width=\linewidth]%
        {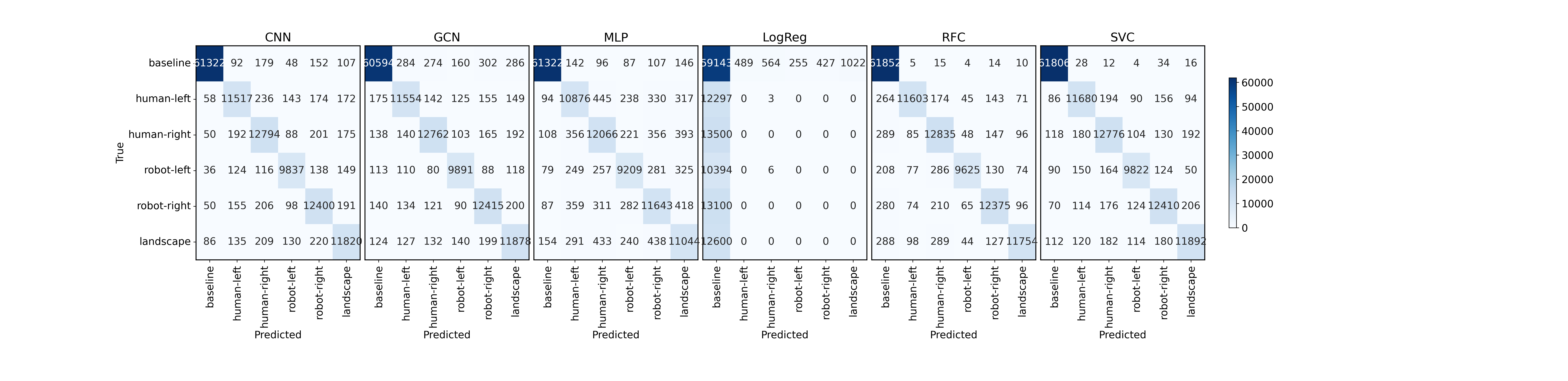}%
    % \end{adjustbox}
    \vspace{-1em}
    \caption{}
  \end{subfigure}%

  %------------------- second panel ------------------%
  \begin{subfigure}[b]{\textwidth}
    \centering
    % \begin{adjustbox}{clip,trim=0 0 0.1\linewidth 0}%
      \includegraphics[width=\linewidth]%
        {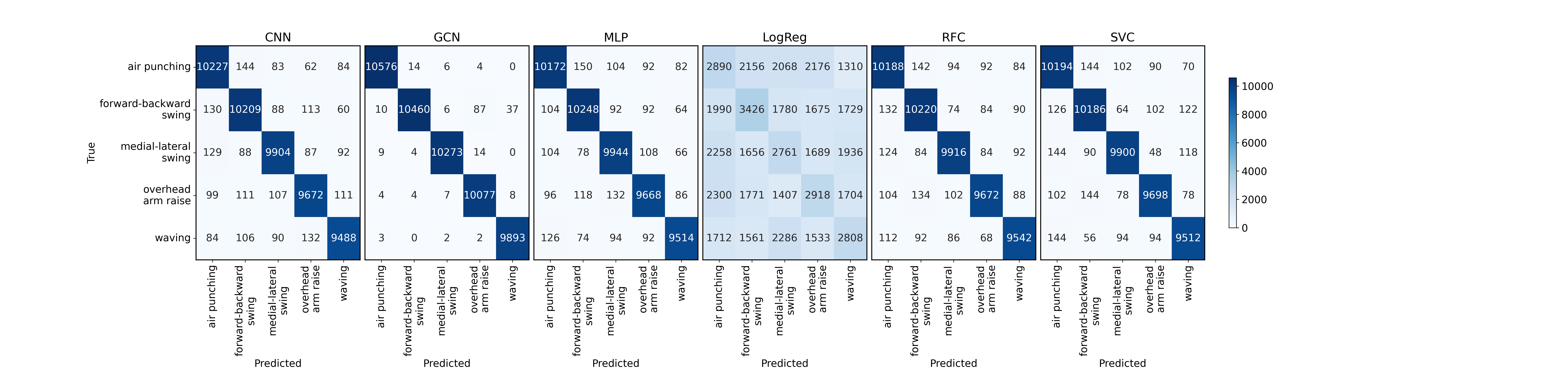}%
    % \end{adjustbox}
    \caption{}
  \end{subfigure}%
  \caption{Confusion‑matrix panels for iCOH in (a) AO‑condition and (b) action‑type tasks.}
  \label{fig:icoh_confmats}
  \vspace{-1em}
\end{figure*}

Figure~\ref{fig:icoh_confmats} shows confusion‐matrix panels for all six classifiers on iCOH. Except for LogReg, models accurately predicted labels in both tasks, though AO‐condition decoding accuracies remain lower due to dataset imbalance. LogReg’s confusion matrices highlight its susceptibility to both class skew and subtle inter‐class distinctions.

% Overall, these results demonstrate that volume‐conduction‐invariant metrics (iCOH, PDC, SpcG) combined with graph‐structured models (GNN) provide the most reliable and generalizable decoding of both AO conditions and observed action types from sensor‐level EEG connectivity.

\section{DISCUSSION}

In this pilot study, EEG data was collected during an action observation (AO) experiment, where healthy participants were exposed to seven different visual conditions and five different types of upper limb actions.
To analyze neural interactions, functional connectivity (FC) was computed for each segment across all six conditions using five widely used FC methods: Coherence (COH), Imaginary Coherence (iCOH), Partial Directed Coherence (PDC), Phase Locking Value (PLV), and Spectral Granger Causality (SpcG). These FC measures were then used to train three machine learning models (Support Vector Machine (SVM), Logistic Regression (LogReg), and Random Forest Classifier (RF)) and three neural-network-based models (Graph Neural Network (GNN), Convolutional Neural Network (CNN), and Multi-Layer Perceptron (MLP) to assess their capabilities in classifying different conditions of AO stimuli and different actions observed in the AO experiment.

Most EEG studies of action observation have typically focused on a single binary classification task—for example, distinguishing AO from baseline or from non‐AO visual controls—which limits insights into how methodological choices generalize across different decoding problems \cite{coll2017crossmodal, ge2019neural}.
By contrast, we evaluated two complementary label sets (AO–condition and action–type), revealing that directed and volume‐conduction–invariant FC metrics exhibited consistently high discriminative power across both tasks. This approach extends earlier work that demonstrated robust decoding of mirror‐neuron engagement in AO‐versus‐baseline paradigms \cite{avanzini2012dynamics, zhangActivationMirrorNeuron2018}, and complements studies of action–type decoding that have applied novel feature extraction but lacked cross‐task validation \cite{xiong2020classifying}.

The consistent performance of metrics such as iCOH, PDC, and spectral Granger causality across decoding problems underscores their potential utility as generalizable biomarkers of motor‐related network engagement \cite{chiarion2023connectivity}.
Our findings suggest that these FC measures capture fundamental properties of mirror‐neuron–mediated processing, supporting their application in translational settings such as stroke rehabilitation via action observation therapy \cite{boni_s_action_2023, coccoOptimizedEEGbasedAction2023}.
Moreover, demonstrating separation of discrete action types informs the design of brain–computer interfaces and neurofeedback protocols that require fine‐grained decoding of intended movements as opposed to simple AO detection \cite{gupta2021subject}. Future work should investigate the stability of these metrics across larger, more diverse cohorts and explore their integration with source‐imaged data to enhance spatial specificity.

% The results of both decoding tasks indicate that iCOH produced the highest number of best-performing models, followed by SpcG and PDC that removing zero-phase-lag interactions enhances the detection of true neural coupling. The success of SpcG and PDC indicates that directed connectivity measures may be more effective in capturing meaningful neural interactions compared to non-directed methods like COH and PLV. 
% Given that iCOH, SpcG, and PDC are all volume conduction-invariant, their superior performance highlights the importance of controlling for volume conduction when analyzing EEG functional connectivity. 

Among the classifiers evaluated, LogReg had the weakest performance, suggesting that LogReg’s assumption of linear relationships limits its ability to capture the intricate, nonlinear dynamics of EEG FC patterns \cite{chiarion2023connectivity}.
In contrast, random forests (RF) strike an effective balance between computational efficiency and accuracy: by averaging across an ensemble of decision trees, RF captures nonlinear feature interactions without extensive hyperparameter tuning, converges rapidly during training, and employs embedded feature‐importance measures to highlight the most discriminative connections \cite{edla2018classification}.
CNN, on the other hand, excels because it learns spatial dependencies within FC adjacency matrices while maintaining translation invariance \cite{wang2019convolutional}.  Its ability to learn data-driven representations allows for greater generalization across FC metrics compared to SVM, which depends on predefined kernels \cite{antony2022classification}.

% GNNs further outperformed all other classifiers, consistently achieving high AUC across all FC measures. They are particularly well suited for EEG‐based FC analysis because they operate directly on graph‐structured data. Whereas CNNs treat the adjacency matrix as an image, GNNs aggregate information from neighboring nodes in a biologically plausible manner, modeling both local and global neural interactions more effectively. Notably, even for FC metrics such as COH and PLV—which yield denser, noisier connectivity estimates and low discrimination for most models—the GNN’s message‐passing framework amplifies informative topological signatures and attenuates noise, uncovering subtle inter‐regional patterns that scalar features alone cannot capture. Furthermore, GNNs integrate both the FC‐based adjacency matrix and node feature matrix (PSD values) \cite{kleplEEGBasedGraphNeural2022}, and this dual‐input mechanism likely contributes to their superior classification performance.  

GNNs delivered the strongest and most consistent performance across all FC metrics, outpacing both CNNs and multilayer perceptrons (MLPs) in convergence speed during training—even on large graph inputs—while modeling neural interactions in a biologically plausible fashion \cite{grana2023review}.
Unlike CNNs, which treat the adjacency matrix as an image, GNNs iteratively aggregate information from node neighborhoods, inherently denoising dense metrics (e.g., COH, PLV) and amplifying subtle topological features that scalar or grid‐based methods may overlook.
A key advantage of our GNN implementation is its dual‐input architecture: by integrating both the FC adjacency matrix and per‐node power spectral density (PSD) features, it leverages complementary views of brain dynamics to achieve superior discrimination—a strategy corroborated in recent EEG‐GNN studies \cite{kleplEEGBasedGraphNeural2022, klepl2024graph}. However, this benefit comes at the cost of requiring additional preprocessing to compute node‐level PSD, a trade‐off that may impact real‐time applicability but is justified by the marked gains in robustness and accuracy.

The extremely high accuracies observed—particularly with GNNs and directed metrics—should be interpreted with caution. Our dataset contained a relatively small number of participants (n = 5), and while stratified cross-validation, early stopping, and multiple repetitions were used to mitigate overfitting, trial-level classification may still inflate apparent performance. In particular, within-subject trial repetition increases the effective training sample size, which can favor deep models such as GNNs that excel at capturing subtle within-subject patterns. The finding that GNNs outperform other classifiers even on weaker metrics (COH, PLV) may reflect their ability to denoise dense connectivity matrices by leveraging graph topology, but it also underscores the need for future replication on larger and more heterogeneous samples. By acknowledging these limitations, we emphasize that these results are best viewed as pilot evidence for the feasibility of FC-based AO decoding rather than definitive proof of model generalizability.

% \begin{figure}
%     \includegraphics[width=\columnwidth]{figures/heatmap_examples_1.png}
%     \caption{Examples of heatmaps of the adjacency matrices of the iCOH metric for all classes. Each cell represents the connectivity strength between a pair of electrodes.}
%     \label{fig:heatmaps}
% \end{figure}

Beyond technical performance, the physiological interpretation of each FC metric provides insight into their relative utility. COH and PLV capture phase synchrony but are confounded by zero-lag interactions, limiting their specificity to genuine neural coupling. By contrast, iCOH explicitly suppresses volume conduction, better isolating long-range, biologically meaningful connections that are critical for distributed motor networks. Directed metrics such as PDC and SpcG model causal influences between regions, consistent with theories of predictive coding in AO, where premotor and parietal areas exert top-down influence on visual processing streams \cite{urgen2020predictive}. The robust performance of iCOH and PDC in particular suggests that AO engages not only synchronous oscillatory networks but also directional information transfer within the sensorimotor system.

%%%%%%%%%%%% LIMITATIONS
Although our results demonstrate strong decoding performance, this pilot investigation’s small sample size (n \(=\) 5) inherently limits the external validity of our findings and elevates the risk of overfitting. Despite using stratified k-fold cross-validation and multiple repetitions to maximize data utilization, a larger and more heterogeneous participant cohort is essential to confirm that these models generalize across different populations and recording conditions. Furthermore, the uneven trial distribution—with baseline trials markedly outnumbering AO conditions—may have biased classifiers toward the more prevalent class. Although we mitigated this skew by reporting macro-averaged metrics (AUC, precision, recall, F1) and balanced accuracy, future studies should pursue targeted data-augmentation, adaptive sampling, or prospectively balanced trial designs to eliminate class imbalances and produce more robust performance estimates.
Our reliance on sensor-level EEG connectivity also omits precise cortical source localization. Future work should employ source localization methods such as dipole modeling and explore efficient model-compression strategies (e.g., pruning, quantization) or lightweight architectures to support low-latency, clinically deployable, and accessible EEG-decoding systems.

\section{CONCLUSION}
This pilot study demonstrates the feasibility of decoding different AO stimuli from sensor-level EEG functional-connectivity patterns, achieving consistently high macro-averaged AUC and balanced accuracy scores across cross-validation folds. By leveraging convolutional neural networks trained on FC adjacency matrices, we provide new evidence that distributed EEG connectivity carries rich information about observed motor actions. 
Although our findings are tempered by the small sample size, trial imbalance, and potential overfitting risks, they highlight the feasibility of decoding both AO conditions and discrete action types from EEG connectivity. This pilot study provides critical guidance for future work: volume-conduction–invariant and directed FC metrics appear to capture physiologically meaningful aspects of AO processing, while GNNs show promise for exploiting graph-structured EEG data. By explicitly linking these computational findings to neurophysiological mechanisms, our results suggest that FC-based decoding strategies may serve as candidate biomarkers for clinical translation, including the optimization of action observation therapy for motor rehabilitation. Expanding to larger and more diverse cohorts, incorporating source-level imaging, and testing in patient populations will be essential next steps to validate these preliminary insights and establish robust, clinically deployable neurotechnologies.

\section*{ACKNOWLEDGMENT}
We thank the National Institutes of Health for funding this research (grant F31-HD102165) and acknowledge Michael Sobrepera and Ajay Anand for their contributions to the robot system development.
%%%%%%%%%%%%%%%%%%%%%%%%%%%%%%%%%%%%%%%%%%%%%%%%%%%%%%%%%%%%%%%%%%%%%%%%%%%%%%%%

\bibliography{refs.bib}

\end{document}